# Ferrimagnetism and spontaneous ordering of transition-metals in $La_2CrFeO_6$ double-perovskite films

S. Chakraverty,[1] A. Ohtomo,[2,*] D. Okuyama,[3] M. Saito,[4] M. Okude,[1] R. Kumai,[5] T. Arima,[6] Y. Tokura,[3,7] S. Tsukimoto,[4] Y. Ikuhara,[4,8] and M. Kawasaki[1,3,4,9]

[1]*Institute for Materials Research, Tohoku University, Sendai 980-8577, Japan*
[2]*Department of Applied Chemistry, Tokyo Institute of Technology, Tokyo 152-8552, Japan*
[3]*Cross-Correlated Materials Research Group, RIKEN Advanced Science Institute, Wako 351-0198, Japan*
[4]*WPI-Advanced Institute for Materials Research, Tohoku University, Sendai 980-8577, Japan*
[5]*National Institute of Advanced Industrial Science and Technology, Tsukuba 305-8562, Japan*
[6]*Institute of Multidisciplinary Research for Advanced Materials, Tohoku University, Sendai 980-8577, Japan*
[7]*Department of Applied Physics, University of Tokyo, Tokyo 113-8656, Japan*
[8]*Institute of Engineering Innovation, University of Tokyo, Tokyo 113-8656, Japan and Nanostructures Research Laboratory, Japan Fine Ceramics Center, Nagoya 456-8587, Japan*
[9]*CREST, Japan Science and Technology Agency, Tokyo 102-0075, Japan*

(Dated: 7/29/2011)

**Abstract**

We report on atomic ordering of B-site transition-metals and magnetic properties of epitaxial $La_2CrFeO_6$ double-perovskite films grown by pulsed-laser deposition under various conditions. The highest ordered sample exhibited a fraction of antisite-disorder of only 0.05 and a saturation magnetization of $\sim 2\mu_B$ per formula unit at 5 K. The result is consistent with the antiferromagnetic ordering of local spin moment ($3d^3_\downarrow 3d^5_\uparrow$; $S = -3/2+5/2 = 1$). Therefore, the magnetic ground state of $La_2CrFeO_6$ double-perovskite that has been long debate is unambiguously revealed to be ferrimagnetic. Our results present a wide opportunity to explore novel magnetic properties of binary transition-metal perovskites upon epitaxial stabilization of the ordered phase.

---

[*] Electronic mail: aohtomo@apc.titech.ac.jp

# I. INTRODUCTION

Complex oxides exhibiting high-Curie temperature ($T_C$) ferromagnetism have been attracting renewed attention for various spin-coupled device applications.[1] When $3d$ and $4d$ or $5d$ transition-metals are combined in a perovskite oxide, they tend to occupy the octahedral site alternately along the [111] direction to form a NaCl-type lattice. This so-called double-perovskite structure is expressed as $A_2M'M''O_6$ where $A$ is an alkaline- or rare-earth element and $M'$ and $M''$ are different transition-metal elements.[2] $Sr_2Fe^{3+}Mo^{5+}O_6$ and $Sr_2Cr^{3+}Re^{5+}O_6$ are well-known examples owing to their half-metallic nature as well as exceptionally high $T_C$.[3,4] For such ideal double-perovskites, a large difference in the formal valence (FV) permits spontaneous ordering of transition-metal elements, thus facile to synthesize in a bulk form. Among the $3d$-$3d$ combinations, however, only a few are known to form double-perovskites such as $La_2Mn^{4+}M''O_6$ ($M'' = Co^{2+}$, $Ni^{2+}$, $Fe^{2+}$)[5,6] where difference in the ionic radius as well as FV is exceptionally large.

Although the spontaneous ordering is not expected for the combination of isovalent $3d$-$3d$ ions,[2] $La_2CrFeO_6$ (LCFO) with $Cr^{3+}/Fe^{3+}$ ($3d^3/3d^5$) has been intensively studied to examine possible ferromagnetism through the $3d^3$-$3d^5$ superexchange according to the Kanamori-Goodenough rule.[7,8] In recent years, a number of theoretical and experimental studies on its ordered form have been reported.[9-13] Based on the local-spin-density calculation, Pickett *et al.* showed that ferrimagnetic ground state with a net spin moment of $2\mu_B$/f.u. (a formula unit) is more stable than ferromagnetic one with ~$7\mu_B$/f.u.[9] On the other hand, Ueda *et al.* postulated that their (111) oriented films, grown by pulsed-laser deposition (PLD) technique in a fashion of the $LaCrO_3/LaFeO_3$ superlattice, exhibited ferromagnetism though the measured saturation magnetization is much less than the expected value.[10,11] Until now, there is no critical evidence to support either scenario.[12,13]

In this communication, we report epitaxial synthesis of LCFO double-perovskite films without using the artificial superlattice technique and their structural and magnetic properties as a function of degree of the Cr/Fe order. Contrary to the common expectation, well-ordered phase is obtained from a single target by the PLD growth under a wide range of growth temperature ($T_g$) and oxygen partial pressure ($P_{O_2}$). The highest ordered sample exhibits the degree of order ~90% and a saturation magnetization ($M_s$) of $2.0 \pm 0.15\mu_B$/f.u. at 5 K. Therefore, the ground-state magnetic order of this compound has been unambiguously verified to be *ferrimagnetic*. The $M_s$ is found to depend on oxidation state as well as the Cr/Fe order, while $T_C$ is nearly constant around 45 ~ 50 K.

# II. EXPERIMENT

Approximately 60 nm-thick LCFO films were grown on atomically flat (111) $SrTiO_3$ substrates using a PLD system with KrF excimer laser pulses (5 Hz) focused on a target (a $LaCr_{0.5}Fe_{0.5}O_3$ disordered ceramic tablet, 99.99% purity) at a fluence of 1.1 J cm$^{-2}$ (laser spot 0.35 x 0.10 cm$^2$).[14] The growth was performed with *in-situ* monitoring of reflection high-energy electron diffraction (RHEED) pattern under conditions shown in Fig. 1 (a). After the growth, samples were quenched to room temperature, keeping $P_{O_2}$ constant. Some of the samples were furnace-annealed in air at 400°C or 800°C for 3 h to refill residual oxygen vacancies in films as well as substrates.

The film composition was analyzed for those grown on (001) MgO substrates under the same conditions by an electron probe microanalyzer (JED-2300F/JSM-6701F, JEOL). The composition of cations was confirmed to be identical to that of the target regardless of growth condition. Surface and crystalline structure of the as-grown films was characterized at room temperature in air by using atomic force microscopy (SPI-400, SII NanoTechnology) and four-circle x-ray diffraction (X'pert MRD, PANalytical) with CuK$\alpha$ radiation ($\lambda$ = 1.541838 Å), respectively.

X-ray diffraction measurements were also performed using a four-circle diffractometer at the synchrotron beamline BL-3A on the Photon Factory, KEK, Japan. The photon energy of the incident x-ray was tuned at 12 keV. All diffraction measurements were carried out at room temperature. The x-ray diffraction intensity of the LCFO was corrected at $\left(\frac{1}{2}\frac{1}{2}\frac{1}{2}\right)$, (1 1 1), $\left(\frac{3}{2}\frac{3}{2}\frac{3}{2}\right)$, (2 2 2), $\left(\frac{5}{2}\frac{5}{2}\frac{5}{2}\right)$, and (3 3 3) reciprocal lattice points. The integrated intensity was estimated by measuring rocking curves along the $\omega$-axis (perpendicular to the scattering plane).

The microstructures and interface structures of the samples were observed using the atomically resolved HAADF-STEM (JEM-2100F, JEOL) equipped with an aberration corrector (CEOS Gmbh) and electron energy loss spectroscopy (GIF Tridiem, Gatan), operating at an acceleration voltage of 200 kV. Moreover, the crystallographic property of the LCFO film was characterized by analyzing selected area diffraction patterns obtained using a conventional 200 kV-TEM (JEM-2010F, JEOL). The electron-transparent thin foil specimens for the TEM and STEM observations were prepared by the common procedures of cutting, mechanically grinding, and dimpling followed by Ar ion milling process (accelerating voltage of 2-4 kV and incidence angle of 6°) to reduce thickness down to several nanometers.

Hysteresis and field-cooled magnetization measurements were performed using a SQUID magnetometer (MPMS-XL, Quantum Design). The hysteresis measurements were done in the field range ± 1 T at a temperature 5 K and the FC measurements were done over the temperature ranging from 5 K to 300 K under a magnetic field of 0.1 T. For clarity, diamagnetic signal of the STO substrates was subtracted.

## III. RESULTS AND DISCUSSION

### A. Thin film growth

Strong growth-condition dependence of the degree of order has been found [Fig. 1 (a)]. Highly ordered films were reproducibly obtained at high-$T_g$-$P_{O_2}$ region and the Cr/Fe order reached up to 90% in the films grown at $T_g$ ~ 1000°C and $P_{O_2}$ = 1 x 10$^{-4}$ Torr (condition A). Moreover, film crystallinity gradually degraded in going away from the condition A [Fig. 1 (b)]. This tendency was also observed in the initial growth. Temporal variations of RHEED intensity is shown in Figs. 1 (c) and (d) for two representative samples grown under conditions A and D, respectively. In the former case, clear oscillation was observed reflecting a layer-by-layer growth mode. As a result, atomically flat surface with monolayer-high (~0.23 nm) steps was seen (inset). In the latter case, intensity damped quickly indicating a three-dimensional growth mode. Correspondingly, film surface became rough exhibiting lateral facets and domain boundaries.

### B. Structural characterizations

In order to evaluate the degree of order, we calculate the x-ray diffraction intensity by using simple crystal model with transition-metal-site disorder and lattice distortion along the [111] direction as shown in Fig. 2. Hereafter we wish to employ the notion of antisite-disorder (*AS*) instead of the Cr/Fe order, which is defined as the percentage of misplaced Cr at Fe site and vice versa.[16] From the model with the AS fraction and atomic displacement of La and O ions towards Fe-rich transition-metal-site plane ($\delta_{La}$ and $\delta_{O}$ in the unit of Å), we calculated the structure factor at each reciprocal lattice point as follows:

$$F_{(\frac{1}{2}\frac{1}{2}\frac{1}{2})} = (1-2AS)(f_{Cr}-f_{Fe}) + 2f_{La}\sin(3\pi\delta_{La}) + 6f_{O}\sin(3\pi\delta_{O}), \tag{1}$$

$$F_{(111)} = f_{Cr} + f_{Fe} - 2f_{La}\sin(6\pi\delta_{La}) - 6f_{O}\sin(6\pi\delta_{O}), \tag{2}$$

$$F_{(\frac{3}{2}\frac{3}{2}\frac{3}{2})} = (1-2AS)(f_{Cr}-f_{Fe}) + 2f_{La}\sin(9\pi\delta_{La}) + 6f_{O}\sin(9\pi\delta_{O}), \tag{3}$$

$$F_{(222)} = f_{Cr} + f_{Fe} - 2f_{La}\sin(12\pi\delta_{La}) - 6f_{O}\sin(12\pi\delta_{O}), \tag{4}$$

$$F_{(\frac{5}{2}\frac{5}{2}\frac{5}{2})} = (1-2AS)(f_{Cr}-f_{Fe}) + 2f_{La}\sin(15\pi\delta_{La}) + 6f_{O}\sin(15\pi\delta_{O}), \tag{5}$$

$$F_{(333)} = f_{Cr} + f_{Fe} - f_{La}\sin(18\pi\delta_{La}) + f_{La}\sin(9\pi(1+2\delta_{La})) \\ - 3f_{O}\sin(18\pi\delta_{O}) + 3f_{O}\sin(9\pi(1+2\delta_{O})). \tag{6}$$

Here, $f_{Cr}$, $f_{Fe}$, $f_{La}$, $f_{O}$, and $F_{(hhh)}$ represent atomic form factors of Cr, Fe, La, O, and structure factor of (*h h h*) reflection, respectively. The calculated intensity *I* was obtained from $I = FF^{*} \times L \times p \times N$. *L*, *p*, and *N* are Lorentz factor, polarization, factor, and scale factor, respectively. The experimentally obtained integrated intensity was fitted by this calculated intensity. The results are given in Figs 3 (a), (b) and (c) for samples grown under conditions A, B and D, respectively. The best results were obtained when (*AS*, $\delta_{La}$, $\delta_{O}$) = (0.051, 0.029, -0.020), (0.145, 0.030, -0.035), and (0.34, -0.038, 0.108) for the samples shown in Figs. 3 (a), (b) and (c), respectively. Apparent magnitude of half-integer reflections with respect to integer reflections systematically decreases with increasing *AS* fraction. It should be mentioned that we constructed the model structures so that all of the half-integer reflections yielded mutual agreements. This is because intensity of the integer reflections has some ambiguity due to insufficient separation from the substrate reflections. As shown in Fig. 3 (d), sharp x-ray rocking curves confirm good crystallinity of the sample grown under condition A. From the (1 1 1) and $(\frac{1}{2} \frac{1}{2} \frac{1}{2})$ reflections, lateral-coherence length of fundamental perovskite lattice and the ordered phase is evaluated to be 250 nm and 220 nm, respectively. As for samples grown under condition B and D, each value is 210 nm/65 nm and 180 nm/40 nm, respectively.

Theoretical aspects of LCFO double-perovskite have been studied in terms of subtle competition between ferromagnetic and antiferromagnetic coupling of local spin moments, to which distance between the nearest-neighbor Fe and Cr ions ($d_{\text{Fe-Cr}}$) is considered in connection with experimental results.[12] Thus, it will be important to determine lattice parameters for our samples. Judging from the position of asymmetric reflections, we have concluded that all the films investigated in this study were pseudomorphically grown on (111) STO [Fig. 3 (e) for example]. From this fact, average distance between the transition-metal ions (*i.e.*, B-O-B along the {100} direction) is deduced through $d_{\text{Fe-Cr}} = \sqrt{d_{111}^2 + (2d_{11\text{-}2})^2}$ where $d_{111}$ is the out-of-plane spacing for LCFO and $d_{11\text{-}2}$ is the in-plane spacing for STO (1.594 Å). As shown in Fig. 3 (f), $d_{\text{Fe-Cr}}$ distributes over *AS* fraction in the intermediate range between those of a powder sample (3.9073 Å; $AS = 0.5$)[16] and superlattice films (3.922 ~ 3.926 Å, assuming the coherent growth on STO; *AS* is unknown).[10,13] Miura *et al.* have found that ground energy minima exists at this range of $d_{\text{Fe-Cr}}$ in both cases of ferromagnetic and antiferromagnetic states.[12] Since we carried out magnetic measurements on annealed samples as well, it is worth mentioning that the lattice parameters remained intact after the annealing at 400°C.

Our results described so far present striking demonstration of spontaneous ordering of Cr and Fe in LCFO films, which is unexpected taking similar ionic radius of Cr and Fe into account ($r_{\text{Cr}^{3+}} = 0.615$ Å and $r_{\text{Fe}^{3+}\text{(HS)}} = 0.645$ Å in an octahedral coordination). In our growth conditions, FV of Cr is expected to be fixed 3+, while that of Fe can be either 2+ or 3+. In Fig. 1 (a), variation in degree of order does not coincide with the difference in FV of Fe and Cr, which is indicated by shaded regions with phase equilibrium lines of simple ferrates. Although actual phase equilibrium of Fe ions must be drawn lower considering the stronger crystal field in perovskite lattice, this contrast implies that tuning FV of Fe is probably not critical for enhancement of the ordering.[17] Although the mechanism of spontaneous ordering is unclear at the moment, we anticipate that lattice disorder has to be reduced to obtain higher Cr/Fe ordering. In fact, variations in degree of order and in rocking curve width show close correlation [Figs. 1 (a) and (b)]. We note that $T_g$ applied for the growth of superlattices (600 ~ 700°C)[10,13] was much lower than $T_g$ applied here. The lattice mismatch to substrate is a deleterious factor because the Cr/Fe ordering as well as film crystallinity became worse when we used (111) LaAlO$_3$ substrates.

Now we would like to discuss the microstructures of our samples. As shown in Fig. 4, atomically abrupt and coherent interface was clearly observed for a sample grown under condition B (*AS* ~ 0.12), indicating formation of elastically distorted region near the interface. Moreover, a SAD pattern into the LCFO layer showed visible (*h h h*) and (*h h k*) diffraction spots with half-integer *h* and *k*, indicating the NaCl-type ordering of Cr and Fe. In addition, the valence state of Cr and Fe ions was confirmed to be both 3+ by the electron energy loss spectroscopy analysis. These results are consistent with the x-ray diffraction analyses. However, we have seen the intensity of the half-integer reflections much weaker near the interface than in bulk region. The disordered region has been found to extend a few nanometers apart from the interface, of which fraction corresponds to at most 5% of the whole sample volume. The presence of interfacial disordered layer is also reflected to the fact that one period of RHEED intensity oscillation corresponded to a charge neutral unit cell (~0.23 nm), but not the double-perovskite unit cell. If the Cr/Fe order had occurred from the beginning, the period would have been twice longer because all deposits must be consumed in order to complete the double-perovskite unit cell. Incomplete surface termination of the STO substrate likely causes formation of the disordered layer, while it is interesting to reveal what triggers the following spontaneous ordering.

### C. Magnetic properties

Using a SQUID magnetometer, magnetization versus applied magnetic field and temperature were measured for samples grown under conditions A to D. For clarity, diamagnetic signal of the STO substrates was subtracted. As shown in Fig. 5 (a), the in-plane magnetization loops taken at 5 K for the samples annealed at 400°C traced well-defined hysteresis, of which magnitude systematically evolved with decreasing $AS$ fraction. As for a sample grown under condition A with $AS = 0.06$, $M_s$ at 1 T was 2.0 $\mu_B$/f.u. with an error arising from inaccuracy of the film volume ($\pm 0.15\mu_B$/f.u.). The magnitude of $M_s$ is consistent with antiferromagnetic ordering of local spin moment ($3d^3_\downarrow 3d^5_\uparrow$; $S = -3/2+5/2 = 1$) and seems remain up to $AS \sim 0.37$. The temperature dependence of field-cooled magnetization showed a clear magnetic transition at $T_C = 45 \sim 50$ K for all the samples [Fig. 5 (b)]. Here we shall emphasis that oxygen annealing led to a drastic enhancement of low-temperature magnetization [Fig. 5 (c)]. In contrast to the annealed samples, as-grown samples indicated broad $M$-$T$ curves without the magnetic transition at $T_C$ (not shown). We believe that suppression of ferrimagnetic ground state is attributed to the presence of residual oxygen vacancies. Other origin like magnetic contamination during the furnace-annealing could be excluded. In fact, no change in $M_s$ was seen for the sample annealed at 800°C (inset).

Finally we briefly discuss the observed systematic dependence of $M_s$ on $AS$ fraction. Considering that LaCrO$_3$ and LaFeO$_3$ show $G$-type antiferromagnetic ordering, the spin moment at the antisite will be aligned antiparallel with respect to the magnetization of the host. In this case, each misplaced Cr ion reduces $M_s$ by $2\mu_B$ ($3d^5_\uparrow \rightarrow 3d^3_\uparrow$; $S = -5/2+2/3 = -1$), while each misplaced Fe ion also reduces $M_s$ by $2\mu_B$ ($3d^3_\downarrow \rightarrow 3d^5_\downarrow$; $S = 3/2-5/2 = -1$), giving a loss of the spin magnetization per antisite of $4\mu_B$. Then, one expects that $M_s$ varies as $(2-4AS)\mu_B$/f.u. and vanishes at $AS = 0.5$. In reality, magnetic frustration arising from relative coordination between misplaced ions would contribute to the deviation of $M_s$.[18]

### IV. CONCLUSION

We have prepared a number of LCFO films on (111) STO substrates at a wide range of growth parameters to study how the ordering as well as physical properties depend on the growth parameters. It has been found that Cr and Fe ions are spontaneously ordered in the NaCl-type double-perovskite structure. Highly ordered phase (up to 90%) could reproducibly be stabilized. Through systematic characterization of the samples having a wide range of the degree of order and lattice parameter, we have unambiguously revealed that magnetic ground state of $La_2CrFeO_6$ double-perovskite is ferrimagnetic.

## ACKNOWLEDFMENTS

S.C. is supported by the Global COE program (Materials Integration), Tohoku University and a Grant-in-Aid for Scientific Research (JSPS). A.O. is supported by the GCOE program (Chemistry), Tokyo Institute of Technology and JSPS. The synchrotron x-ray study was performed with the approval of the Photon Factory Program Advisory Committee (No.2009S2-003). The work was partly supported by JSPS through its "Funding Program for World-Leading Innovative R&D on Science and Technology (FIRST Program)".

# References

[1]  E. Dagotto and Y. Tokura, MRS Bull. **33**, 1037 (2008).

[2]  M. T. Anderson, K. B. Greenwood, G. A. Taylor, and K. R. Poeppelmeier, Prog. Solid State Chem. **22**, 197 (1993).

[3]  K.-I. Kobayashi, T. Kimura, H. Sawada, K. Terakura, and Y. Tokura, Nature **395**, 677 (1998).

[4]  H. Kato, T. Okuda, Y. Okimoto, Y. Tomioka, Y. Takenoya, A. Ohkubo, M. Kawasaki, and Y. Tokura, Appl. Phys. Lett. **81**, 328 (2002).

[5]  G. Blasse, J. Phys. Chem. Solids **26**, 1969 (1965).

[6]  K. Ueda, Y. Muraoka, H. Tabata, and T. Kawai, Appl. Phys. Lett. **78**, 512 (2001).

[7]  J. Kanamori, J. Phys. Chem. Solids **10**, 87 (1959).

[8]  J. B. Goodenough, Phys. Rev. **100**, 564 (1955).

[9]  W. E. Pickett, Phys. Rev. B **57**, 10613 (1998).

[10]  K. Ueda, H. Tabata, and T. Kawai, Science **280**, 1064 (1998).

[11]  W. E. Pickett, G. I. Meijer, K. Ueda, H. Tabata, and T. Kawai, Science **281**, 1571a (1998).

[12]  K. Miura and K. Terakura, Phys. Rev. B **63**, 104402 (2001).

[13]  B. Gray, H. N. Lee, J. Liu, J. Chakhalian, and J. W. Freeland, Appl. Phys. Lett. **97**, 013105 (2010) and references therein.

[14]  S. Chakraverty, A. Ohtomo, M. Okude, K. Ueno, and M. Kawasaki, Cryst. Growth Des. **10**, 1725 (2010).

[15]  The range of the Cr/Fe order or AS fraction shown in Figs. 1 (a), 3 (f), and 5 (c) is determined from intensity ratio of the LCFO $\left(\frac{1}{2}\frac{1}{2}\frac{1}{2}\right)$ reflection to the (1 1 1) reflection, which was taken by using an in-house XRD and calibrated with the data taken by a synchrotron XRD [shown in Figs. 3 (a) to 3 (c)].

[16]  A. K. Azad, A. Mellergård, S.-G. Eriksson, S. A. Ivanov, S. M. Yunus, F. Lindberg, G. Svensson, and R. Mathieu, Mater. Res. Bull. **40**, 1633 (2005).

[17]  T. Manako, M. Izumi, Y. Konishi, K.-I. Kobayashi, M. Kawasaki, and Y. Tokura, Appl. Phys. Lett. **74**, 2215 (1999).

[18]  D. Serrate, J. M. De Teresa, and M. R. Ibarra, J. Phys.: Condens. Matter **19**, 023201 (2007).

**Figures:**

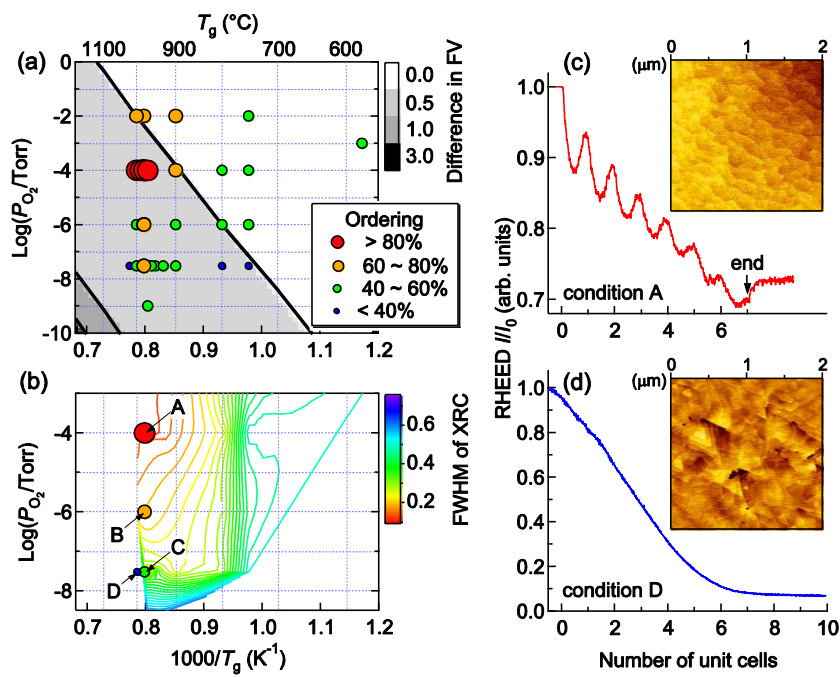

FIG. 1: (Color online) (a) Epitaxial growth conditions of LCFO films mapped in a $T_g$-$P_{O_2}$ diagram. Circles with different size and color represent the growth conditions, where degree of order determined by XRD are > 80% (large and red), 60 ~ 80% (middle and orange), 40 ~ 60% (small and green) and < 20% (the smallest and blue). Solid lines and shaded regions indicate phase equilibrium of Fe-O binary system and stable conditions for coexistence of $Cr_2O_3$ and Fe, FeO, $Fe_3O_4$, or $Fe_2O_3$ (ordered from lower left to upper right), respectively. (b) Contour mapping of the FWHM of rocking curves for the LCFO (1 1 1) reflection. (c) and (d) RHEED intensity oscillations for the specular beam during the initial growth of LCFO films on STO (111) substrates under conditions A and D, respectively. The insets depict AFM images of each film. The range of color code corresponds to ~3 nm in height.

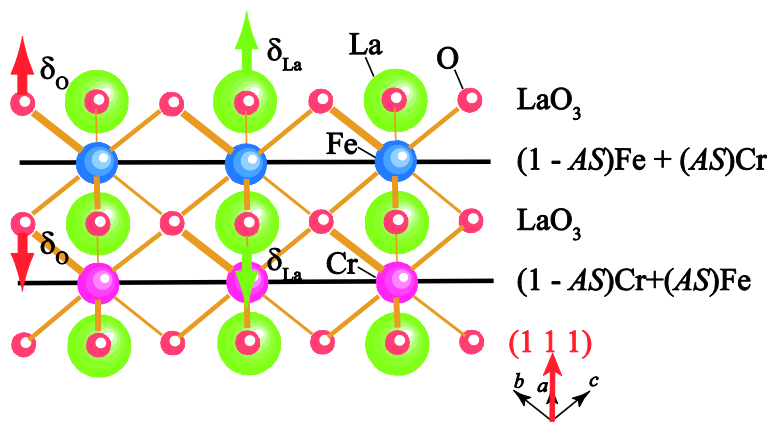

FIG. 2: (Color online) Schematic crystal-structure-model of $La_2CrFeO_6$. In this model, the atomic displacement of La and O are indicated by the $\delta_{La}$ and $\delta_O$. $AS$ is the degree of antisite-disorder.

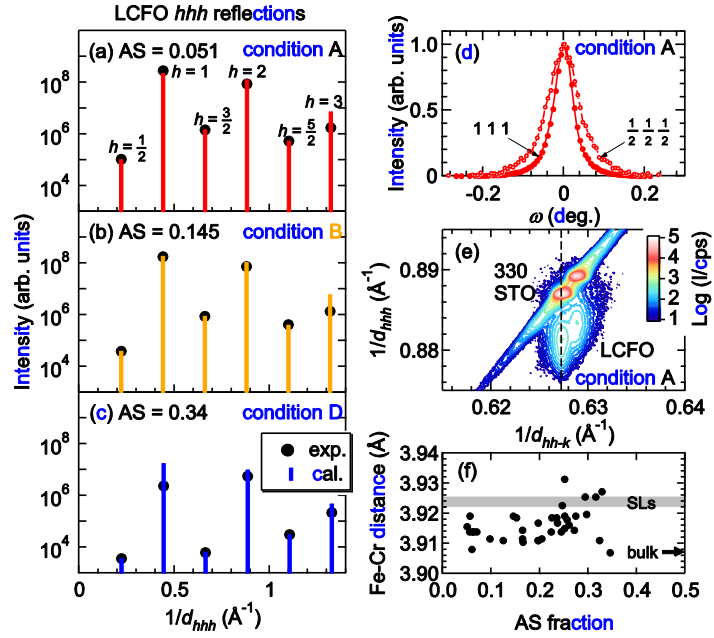

FIG. 3: (Color online) (a-c) X-ray diffraction profiles of the samples grown under conditions A, B and D, respectively. The experimentally observed intensity and calculated one are indicated by circles and bars, respectively. (d) X-ray rocking curves of LCFO (1 1 1) and $\left(\frac{1}{2}\frac{1}{2}\frac{1}{2}\right)$ reflections taken for sample grown under condition A. (e) X-ray reciprocal space mapping around STO (3 3 0) reflection taken for a sample grown under condition A. Double reflections are due to CuK$\alpha_1$ and CuK$\alpha_2$ radiations. (f) Distance between the nearest-neighbor Fe and Cr ions as a function of antisite- disorder (*AS*) fraction. SLs and bulk are referred as to values reported for artificial superlattices[10,13] and a powder sample with $AS = 0.5$.[16]

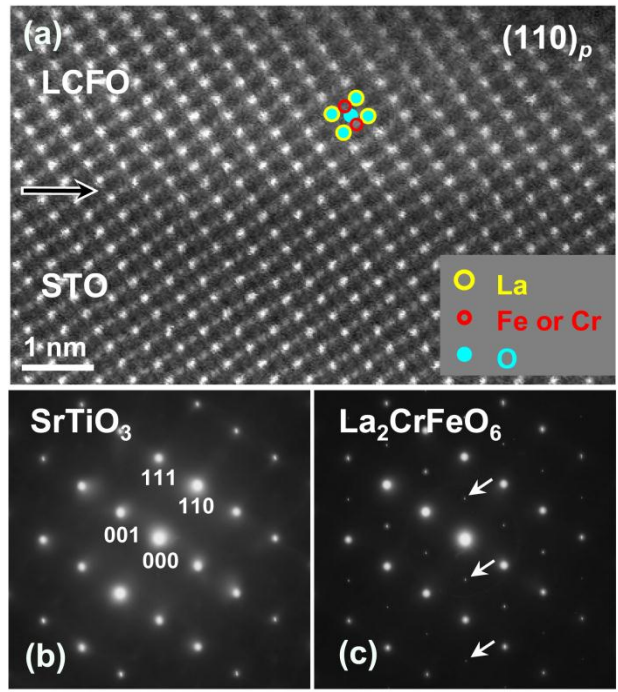

FIG. 4: (Color online) (a) HAADF-STEM image observed along a substrate [110] zone axis for a sample grown under condition B. Circles highlight positions of La (bright spots), Fe/Cr (weak spots), and O (background), respectively. Horizontal arrow indicates position of the interface between LCFO film and STO substrate. Selected area diffraction patterns of STO (b) and LCFO (c). The presence of ($h\,h\,h$) and ($h\,h\,k$) diffraction spots with half-integer $h$ and $k$ in the latter indicates the NaCl-type ordering of B-site cations.

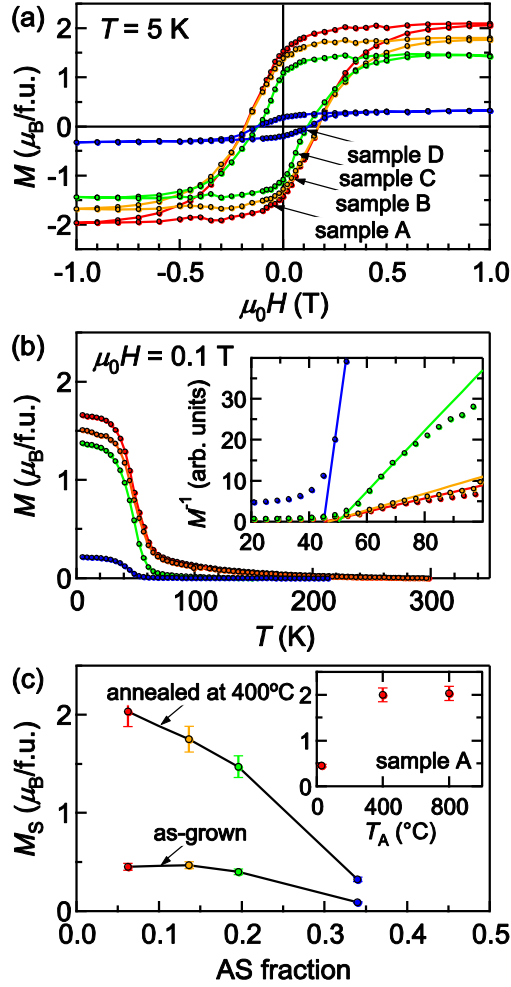

FIG. 5: (Color online) (a) Magnetization hysteresis curves taken at 5 K for samples grown under conditions A to D and annealed at 400°C. (b) The temperature dependence of field-cooled magnetization for the same samples taken during warming under 0.1 T. Inset shows the temperature dependence of inverse magnetization. Solid lines are linear fits to the plots above $T_c$. (c) Saturation magnetization ($M_s = M$ at 5 K under 1 T) for as-grown and annealed (400°C) samples as a function of antisite-disorder fraction. Inset depicts the annealing temperature ($T_A$) dependence of $M_s$ for sample A. The error bars reflect inaccuracy of the sample volume.